\documentclass[a4paper,twoside]{article}

\usepackage{epsfig}
\usepackage{subfigure}
\usepackage{calc}
\usepackage{amssymb}
\usepackage{amstext}
\usepackage{amsmath}
\usepackage{amsthm}
\usepackage{multicol}
\usepackage{pslatex}
\usepackage{apalike}
\usepackage{SCITEPRESS}     
\usepackage{bm}
\usepackage{epstopdf}

\newtheorem{idefinition}{Definition}
\newtheorem{itheorem}{Theorem}

\begin{document}

\title{Quantification of De-anonymization Risks in Social Networks}

\author{
\authorname{Wei-Han Lee\sup{1}, Changchang Liu\sup{1}, Shouling Ji\sup{2}, Prateek Mittal\sup{1} and Ruby Lee\sup{1}}
\affiliation{\sup{1}Princeton University, Princeton, U.S.A}
\affiliation{\sup{2} Zhejiang University and Georgia Institute of Technology, Atlanta, U.S.A }
\email{\{weihanl, cl12, pmittal, rblee\}@princeton.edu, sji@gatech.edu}
}


\abstract{The risks of publishing privacy-sensitive data have received considerable attention recently.
Several de-anonymization attacks have been proposed to re-identify individuals even if data anonymization techniques were applied. However, there is no theoretical quantification for relating the data utility that is preserved by the anonymization techniques and the data vulnerability against de-anonymization attacks.\\
~~~~~~In this paper, we theoretically analyze the de-anonymization attacks and provide conditions on the utility of the anonymized data (denoted by anonymized utility) to achieve successful de-anonymization. To the best of our knowledge, this is the first work on quantifying the relationships between anonymized utility and de-anonymization capability. Unlike previous work, our quantification analysis requires no assumptions about the graph model, thus providing a general theoretical guide for developing practical de-anonymization/anonymization techniques. \\
~~~~~~Furthermore, we evaluate state-of-the-art de-anonymization attacks on a real-world Facebook dataset to show the limitations of previous work. By comparing these experimental results and the theoretically achievable de-anonymization capability derived in our analysis, we further demonstrate the ineffectiveness of previous de-anonymization attacks and the potential of more powerful de-anonymization attacks in the future.}

\onecolumn \maketitle \normalsize \vfill
\section{\uppercase{Introduction}}
\label{sec:introduction}
Individual users' data such as social relationships, medical records and mobility traces are becoming increasingly important for application developers and data-mining researchers. These data usually contain sensitive and private information about users. Therefore, several data anonymization techniques have been proposed to protect users' privacy \cite{hay2007anonymizing}, \cite{liu2008SIGMOD}, \cite{Pedarsani2011SIGKDD}.\\
\indent The privacy-sensitive data that are closely related to individual behavior usually contain rich graph structural characteristics. For instance, social network data can be modeled as graphs in a straightforward manner. Mobility traces can also be modeled as graph topologies according to \cite{Srivatsa2012CCS}. Many people nowadays have accounts through various social networks such as Facebook, Twitter, Google+, Myspace and Flicker. Therefore, even equipped with advanced anonymization techniques, the privacy of structural data still suffers from de-anonymization attacks assuming that the adversaries have access to rich auxiliary information from other channels \cite{backstrom2007WWW}, \cite{Narayanan2008SP}, \cite{Narayanan2009SP},  \cite{Srivatsa2012CCS}, \cite{ji2014CCS}, \cite{nilizadeh2014CCS}. Narayanan et al. \cite{Narayanan2009SP} effectively de-anonymized a Twitter dataset by utilizing a Flickr dataset as auxiliary information based on the inherent cross-site correlations. Nilizadeh et al. \cite{nilizadeh2014CCS} exploited the community structure of graphs to de-anonymize social networks. Furthermore, Srivatsa et al. \cite{Srivatsa2012CCS} proposed to de-anonymize a set of location traces based on a social network.\\
\indent However, to the best of our knowledge, there is no work on theoretically quantifying the data anonymization techniques to defend against de-anonymization attacks. In this paper, we aim to theoretically analyze the de-anonymization attacks in order to provide effective guidelines for evaluating the threats of future de-anonymization attacks. We aim to rigorously evaluate the vulnerabilities of existing anonymization techniques. For an anonymization approach, not only the users' sensitive information should be protected, but also the anonymized data should remain useful for applications, i.e., the \textit{anonymized utility} should be guaranteed. Then, under what range of anonymized utility, is it possible for the privacy of an individual to be broken? We will quantify the vulnerabilities of existing anonymization techniques and establish the inherent relationships between the application-specific anonymized utility and the de-anonymization capability.  Our quantification not only provides theoretical foundations for existing de-anonymization attacks, but also can serve as a guide for designing new de-anonymization and anonymization schemes. For example, the comparison between the theoretical de-anonymization capability and the practical experimental results of current de-anonymization attacks demonstrates the ineffectiveness of existing de-anonymization attacks. Overall, we make the following contributions:
\begin{itemize}
\item{We theoretically analyze the performance of structure-based de-anonymization attacks through formally quantifying the vulnerabilities of anonymization techniques. Furthermore, we rigorously quantify the relationships between the de-anonymization capability and the utility of anonymized data, which is the first such attempt to the best of our knowledge. Our quantification provides theoretical foundations for existing structure-based de-anonymization attacks, and can also serve as a guideline for evaluating the effectiveness of new de-anonymization and anonymization schemes through comparing their corresponding de-anonymization performance with our derived theoretical bounds.}
\item{To demonstrate the ineffectiveness of existing de-anonymization attacks, we implemented these attacks on a real-world Facebook dataset. Experimental results show that previous methods are not robust to data perturbations and there is a significant gap between their de-anonymization performance and our derived theoretically achievable de-anonymization capability. This analysis further demonstrates the potential of developing more powerful de-anonymization attacks in the future.}
\end{itemize}

\vspace{-2em}
\section{\uppercase{Related Work}}
\vspace{-1em}
\subsection{Challenges for Anonymization Techniques}
Privacy preservation on structural data has been studied extensively. The naive method is to remove users' personal identities (e.g., names, social security numbers), which, unfortunately, is rather vulnerable to structure-based de-anonymization attacks \cite{backstrom2007WWW}, \cite{Narayanan2008SP}, \cite{Narayanan2009SP}, \cite{hay2008VLDB}, \cite{liu2008SIGMOD}, \cite{Srivatsa2012CCS}, \cite{ji2014CCS}, \cite{sharad2014automated}, \cite{sharad2013anonymizing}, \cite{nilizadeh2014CCS}, \cite{buccafurri2015discovering}. An advanced mechanism, $k$-anonymity, was proposed in \cite{hay2008VLDB}, which obfuscates the attributes of users so that each user is indistinguishable from at least $k-1$ other users. 
Although $k$-anonymity has been well adopted, it still suffers from severe privacy problems due to the lack of diversity with respect to the sensitive attributes as stated in \cite{machanavajjhala:TKDD07}.
 Differential privacy~\cite{dwork2011differential}, \cite{liu2016ddp} is a popular privacy metric that statistically minimizes the privacy leakage. 
Sala et al. in \cite{sala:imc11} proposed to share a graph in a differentially private manner. However, to enable the applicability of such an anonymized graph, the differential private parameter should not be large, which would thus make their method ineffective in defending against structure-based de-anonymization attacks \cite{ji2014CCS}.
Hay et al. in \cite{hay2007anonymizing} proposed a perturbation algorithm that applies a sequence of $r$ edge deletions followed by $r$ other random edge insertions. However, their method also suffers from structure-based de-anonymization attacks as shown in \cite{nilizadeh2014CCS}.\\
\indent In summary, existing anonymization techniques are subject to two intrinsic limitations: 1) they are not scalable and thus would fail on high-dimensional datasets; 2) They are susceptible to adversaries that leverage the rich amount of auxiliary information to achieve structure-based de-anonymization attacks.
\subsection{De-anonymization Techniques}
Structure-based de-anonymization was first introduced in \cite{backstrom2007WWW}, where both active and passive attacks were discussed. However, the limitation of scalability reduces the effectiveness of both attacks.\\
\indent \cite{Narayanan2008SP} utilized the Internet movie database as the source of background knowledge to successfully identify users' Netflix records, uncovering their political preferences and other potentially sensitive information. In \cite{Narayanan2009SP}, the authors further de-anonymized a Twitter dataset using a Flickr dataset as auxiliary information. They proposed the popular seed identification and mapping propagation process for de-anonymization. In order to obtain the seeds, they assume that the attacker has access to a small number of members of the target network and can determine if these members are also present in the auxiliary network (e.g., by matching user names and other contextual information). The authors in \cite{Srivatsa2012CCS} captured the WiFi hotspot and constructed a contact graph by connecting users who are likely to utilize the same WiFi hotspot for a long time. Based on the fact that friends (or people with other social relationships) are likely to appear in the same location, they showed how mobility traces can be de-anonymized using an auxiliary social network. However, their de-anonymization approach is rather time-consuming and may be computationally infeasible for real applications. 
In~\cite{sharad2013anonymizing},~\cite{sharad2014automated}, Sharad et al. studied the de-anonymization attacks on ego graphs with graph radius of one or two and they only studied the linkage of nodes with degree greater than 5. As shown in  previous work~\cite{ji2014CCS}, nodes with degree less than 5 cannot be ignored since they form a large portion of the original real-world data.
Recently, \cite{nilizadeh2014CCS} proposed a community-enhanced de-anonymization scheme for social networks. The community-level de-anonymization is first implemented for finding more seed information, which would be leveraged for improving the overall de-anonymization performance. Their method may, however, suffer from the serious inconsistency problem of community detection algorithms. \\
\indent Most de-anonymization attacks are based on the seed-identification scheme, which {either relies on the adversary's prior knowledge or a seed mapping process. Limited work has been proposed that requires no prior seed knowledge by the adversary \cite{pedarsani2013Allerton},\cite{ji2014CCS}. In \cite{pedarsani2013Allerton}, the authors proposed a Bayesian-inference approach for de-anonymization. However, their method is limited to de-anonymizing sparse graphs. \cite{ji2014CCS} proposed a cold-start optimization-based de-anonymization attack. However, they only utilized very limited structural information (degree, neighborhood, top-K reference distance and sampling closeness centrality) of the graph topologies. \\ 
\indent Ji et al. further made a detailed comparison for the performance of existing de-anonymization techniques in \cite{ji2015secgraph}.
\subsection{Theoretical Work for De-anonymization}
Despite these empirical de-anonymization methods, limited research has provided theoretical analysis for such attacks. The authors in \cite{Pedarsani2011SIGKDD} conducted preliminary analysis for quantifying the privacy of an anonymized graph $G$ according to the ER graph model \cite{erdHos:selected76}. However, their network model (ER model) may not be realistic, since the degree distribution of the ER model (follows the Poisson distribution) is quite different from the degree distributions of most observed real-world structural data  \cite{Newman2010oxford}, \cite{Newman2003SIAM}. \\
\indent Ji et al. in \cite{ji2014CCS} further considered a configuration model to quantify perfect de-anonymization and $(1-\epsilon)$-perfect de-anonymization. However, their configuration model is also not general for many real-world data structures. Furthermore, their assumption that the anonymized and the auxiliary graphs are sampled from a conceptual graph is not practical since only edge deletions from the conceptual graph have been considered. In reality, 
 edge insertions should also be taken into consideration. Besides, neither \cite{Pedarsani2011SIGKDD} nor \cite{ji2014CCS} formally analyzed the relationships between the de-anonymization capability and the anonymization performance (e.g., the utility performance for the anonymization schemes).\\
\indent Note that our theoretical analysis in Section~\ref{sec:quantification} takes the application-specific utility definition into consideration. Such non-linear utility analysis makes the incorporation of edge insertions to our quantification rather nontrivial. Furthermore, our theoretical quantification does not make any restrictive assumptions about the graph model. Therefore, our theoretical analysis would provide an important guide for relating de-anonymization capability and application-specific anonymizing utility.\\
 \indent Further study on de-anonymization attacks can be found in~\cite{fabiana2015anonymizing}, \cite{ji2015your}, \cite{korula2014efficient}. These papers provide {theoretically guaranteed performance bounds} for their de-anonymization algorithms. However, their derived performance bounds can only be guaranteed under restricted assumptions of the random graph, such as ER model and power-law model. We will show the advantage of our analysis over these approaches where our analysis requires no assumptions or constraints on the graph model as these approaches required.

\section{\uppercase{System Model}}
We model the structural data (e.g., social networks, mobility traces, etc.) as a graph, where the nodes represent users who are connected by certain relationships (social relationships, mobility contacts, etc.). The anonymized graph can be modeled as $G_a=(V_a,E_a)$, where $V_a=\{i|i$ is an anonymized node\} is the set of users and $E_a=\{e_a(i,j)|e_a(i,j)$ is the relationship between $i\in V_a$ and $j\in V_a$\} is the set of relationships. Here, $e_a(i,j)=1$ represents the existence of a connecting edge between $i$ and $j$ in $G_a$, and $e_a(i,j)=0$ represents the non-existence of such an edge. The neighborhood of node $i\in V_a$ is $N_a(i)=\{ j| e_a(i,j)=1\}$ and the degree is defined as $|N_a(i)|$.\\
\indent Similarly, the auxiliary structural data can also be modeled as a graph $G_u=(V_u,E_u)$ where $V_u$ is the set of labelled (known) users and $E_u$ is the set of relationships between these users. Note that the auxiliary (background) data can be easily obtained through various channels, e.g., academic data mining, online crawling, advertising and third-party applications \cite{Narayanan2009SP,Pedarsani2011SIGKDD,pham2013SIGMOD,Srivatsa2012CCS}.\\
\indent A de-anonymization process is a mapping $\sigma : V_a \rightarrow V_u$. $\forall i \in V_a$, its mapping under $\sigma$ is $\sigma(i)\in V_u \cup \{\perp\}$, where $\perp$ indicates a non-existent (null) node. Similarly, $\forall e_a(i,j) \in E_a$, $\sigma (e_a(i,j)) = e_u(\sigma (i), \sigma (j)) \in E_u \cup \{\perp\}$. Under $\sigma$, a successful de-anonymization on $i \in V_a$ is defined as $\sigma (i) = i$, if $i \in V_u$ or $\sigma (i) =\perp$, otherwise. For other cases, the de-anonymization on $i$ fails.

\subsection{\uppercase{Attack model}}
We assume that the adversary has access to $G_a = (V_a, E_a)$ and $G_u= (V_u,E_u)$. $G_a=(V_a, E_a)$ is the anonymized graph and the adversary can only get access to the structural information of $G_a$. $G_u=(V_u,E_u)$ is the auxiliary graph and the adversary already knows all the identities of the nodes in $G_u$. In addition, we do not assume that the adversary has other prior information (e.g., seed information). These assumptions are more reasonable than most of the state-of-the-art research \cite{Narayanan2009SP,Srivatsa2012CCS,nilizadeh2014CCS}.

\section{\uppercase{Theoretical Analysis}} \label{sec:quantification}
In this section, we provide a theoretical analysis for the structure-based de-anonymization attacks. 
Under any anonymization technique, the users' sensitive information should be protected without significantly affecting the utility of the anonymized data for real systems or research applications. We aim to quantify the trade-off between preserving users' privacy and the utility of anonymized data. Under what range of anonymized utility, is it possible for the privacy of an individual to be broken (i.e., for the success of de-anonymization attacks)? To answer this, we quantify the limitations of existing anonymization schemes and establish an inherent relationship between the anonymized utility and de-anonymization capability. Our theoretical analysis incorporates an application-specific utility metric for the anonymized graph, which further makes our rigorous quantification useful for real world scenarios.
Our theoretical analysis can serve as an effective guideline for evaluating the performance of practical de-anonymization/anonymization schemes (will be discussed in Section~\ref{tradeoff}).\\
\indent First, we assume that there exists a conceptually underlying graph $G=(V,E)$ with $V=V_a \cup V_u$ and $E$ is a set of relationships among users in $V$, where $e(i,j)=1\in E$ represents the existence of a connecting edge between $i$ and $j$, and $e(i,j)=0\in E$ represents the non-existence of such an edge. Consequently, $G_a$ and $G_u$ could be viewed as observable versions of $G$ by applying edge insertions or deletions on $G$ according to proper relationships, such as `co-occurrence' relationships in Gowalla \cite{pham2013SIGMOD}. In comparison, previous work \cite{ji2014CCS,Pedarsani2011SIGKDD} only considers edge deletions which is an unrealistic assumption.\\
\indent For edge insertions from $G$ to $G_a$, the process is: $\forall e(i,j)=0 \in E$, $e(i,j)=1$ appears in $E_a$ with probability $p_a^{\mathrm{add}}$, i.e., $Pr(e_a(i,j) =1|e(i,j)=0) = p_a^{\mathrm{add}}$. The probability of edge deletion from $G$ to $G_a$ is $p_a^{\mathrm{del}}$, i.e., $Pr(e_a(i,j) =0|e(i,j)=1) = p_a^{\mathrm{del}}$. Similarly, the insertions and deletions from $G$ to $G_u$ can be characterized with probabilities $p_u^{\mathrm{add}}$ and $p_u^{\mathrm{del}}$. Furthermore, we assume that both the insertion/deletion relationship of each edge is independent of every other edge. Furthermore, this model is intuitively reasonable since the three graphs $G$, $G_a$, $G_u$ are related with each other. In addition, our model is more reasonable than the existing models in \cite{ji2014CCS,Pedarsani2011SIGKDD} because we take both edge deletions and insertions into consideration. Note that the incorporation of edge insertion is non-trivial in our quantification of non-linear application-specific utility analysis. Our quantification analysis would therefore contribute to relating the real world application-specific anonymizing utility and the de-anonymization capability.\\
\indent Adjacency matrix and transition probability matrix are two important descriptions of a graph, and the graph utility is also closely related to these matrices. The adjacency matrix is a means of representing which nodes of a graph are adjacent to which other nodes. We denote the adjacency matrix by $\bm{A}$ (resp. ${\bm A}_a$ and ${\bm A}_u$) for graph $G$ (resp. $G_a$ and $G_u$), where the element ${\bm A}(i,j) = e(i,j)$ (resp. ${\bm A}_a(i,j) = e_a(i,j)$ and ${\bm A}_u(i,j) = e_u(i,j)$). Furthermore, the transition probability matrix is a matrix consisting of the one-step transition probabilities, which is the probability of transitioning from one node to another in a single step. We denote the transition probability matrix by ${\bm T}$ (resp. ${\bm T}_a$ and ${\bm T}_u$) for graph $G$ (resp. $G_a$ and $G_u$), where the element ${\bm T}(i,j) = e(i,j)/deg(i)$ (resp. ${\bm T}_a(i,j) = e_a(i,j)/deg_a(i)$ and ${\bm T}_u(i,j) = e_u(i,j)/deg_u(i)$), and $deg(i),deg_a(i),deg_u(i)$ represent the degree of node $i$ in $G,G_a,G_u$, respectively.\\
\indent We now define the smallest ($l$) and largest ($h$) probabilities of an edge existing between two nodes in the graph $G$, and the graph density (denoted by $R$). For graph $G$, we denote $|V|=N$ and $|E|=M$. Let $p(i,j)$ be the probability of an edge existing between $i, j \in V$ and define $l = \min \{p(i,j) |i, j \in V, i  \neq j\}$, $h = \max \{ p(i,j) |i, j \in V, i \neq j\}$, the expected number of edges $ P_{T}=\sum_{i,j\in V} p(i,j)$ and the graph density $R=\frac{P_T}{\binom{N}{2}}$.\\
\indent Then, we start our formal quantification from the simplest scenario where the anonymized data and the auxiliary data correspond to the same group of users i.e., $V_a=V_u$ as in \cite{Narayanan2009SP,Pedarsani2011SIGKDD,Srivatsa2012CCS}. This assumption does not limit our theoretical analysis since we can either (a) apply it to the overlapped users between $V_a$ and $V_u$ or (b) extend the set of users to $V_a^{new} = V_a \cup (V_u \backslash V_a)$ and $V_u^{new} = V_u \cup (V_a \backslash V_u)$, and apply the analysis to $G_a = (V_a^{new}, E_a)$ and $G_u = (V_u^{new}, E_u)$. Therefore, in order to prevent any confusion and without loss of generality, we assume $V_a = V_u$ in our theoretical analysis. We define $\sigma_k$ as a mapping between $G_a$ and $G_u$ that contains $k$ incorrectly-mapped pairs.\\ 
\indent Given a mapping $\sigma: V_a \rightarrow V_u$, we define the \emph{Difference of Common Neighbors (DCN)} on a node $i$'s mapping $\sigma(i)$ as $\phi_{i,\sigma (i)} = |N_a^i \backslash N_u^{\sigma (i)} | + |N_u^{\sigma (i)} \backslash N_a^i |$, which measures the neighborhoods' difference between node $i$ in $G_a$ and node $\sigma (i)$ in $G_u$ under the mapping $\sigma$. Then, we define the overall \emph{DCN} for all the nodes under the mapping $\sigma$ as $\Phi_\sigma =\sum_{(i,\sigma (i))\in \sigma} \phi_{i,\sigma (i)}$.\\
\indent Next, we not only explain why structure-based de-anonymization attacks work but also quantify the trade-off between the anonymized utility and the de-anonymization capability. We first quantify the relationship between a straightforward utility metric, named \emph{local neighborhood utility}, and the de-anonymization capability. Then we carefully analyze a more general utility metric, named \emph{global structure utility}, to accommodate a broad class of real-world applications.

\begin{figure*}[!t]
\centering
\includegraphics[width=6.3in,height=1.9in]{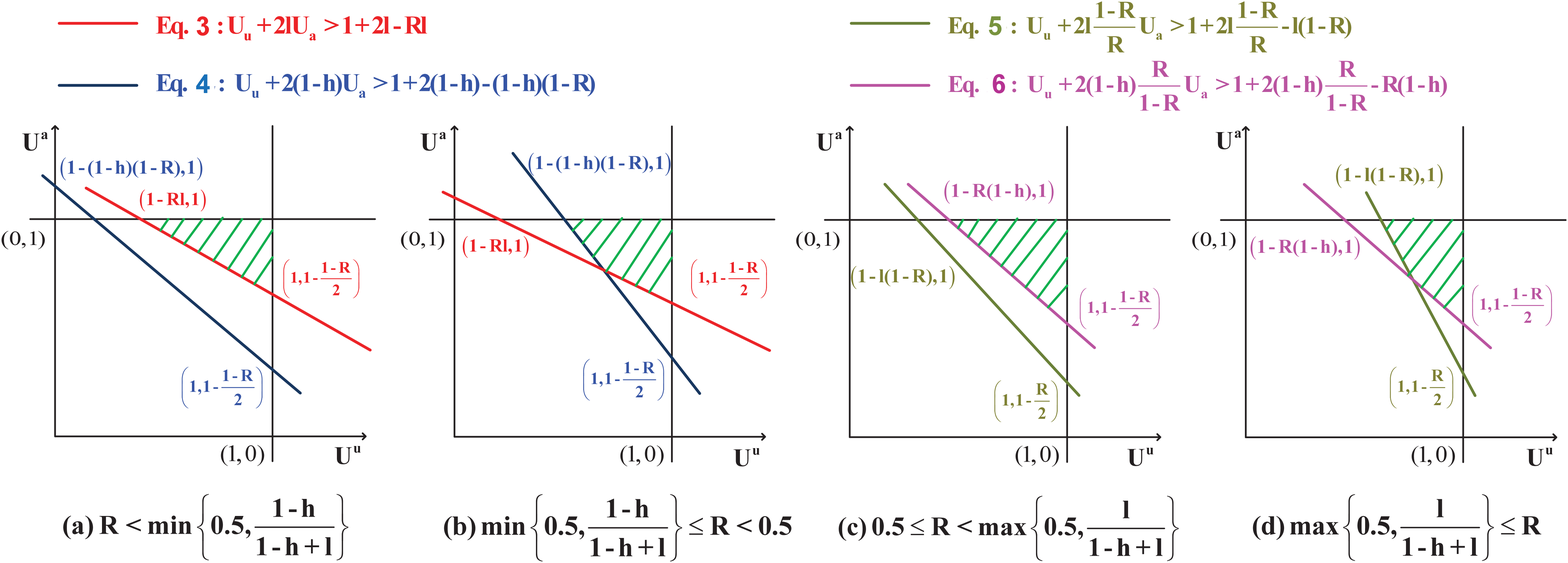}
\caption{Visualization of utility region (green shaded) for successful de-anonymization under different scenarios. To guarantee the applicability of the anonymized data, the anonymized utility should be preserved by the anonymization techniques. We theoretically demonstrate that successful de-anonymization can be achieved if the anonymized utility locates within these shaded regions.}
\label{bound}
\end{figure*}

\subsection{Relation Between the Local Neighborhood Utility and De-anonymization Capability}
At the beginning, we explore a straightforward utility metric, local neighborhood utility, which evaluates the distortion of the anonymized graph $G_a$ from the conceptually underlying graph $G$ as
\begin{idefinition}\label{def_utility} The local neighborhood utility for the anonymized graph is $U_a=1-\frac{||{\bm A}_a-{\bm A}||_1}{N(N-1)}= 1-\frac{\mathbb{E}[D(G_a,G)]}{N(N-1)}$ (the denominator is a normalizing factor to guarantee $U_a\in [0,1]$), where $D(\cdot , \cdot )$ is the hamming distance \cite{hamming1950belljournal} of edges between two graphs, i.e., if $e_a(i,j) \neq e(i,j)$, $D(e_a(i,j),e(i,j))=1$ and $\mathbb{E}[D(G_a,G)]$ is the distortion between $G_a$ and $G$ and $\mathbb{E}[D(G_a,G)] = \mathbb{E}[\sum\limits_{i,j} D(e_a(i,j),e(i,j))]=\sum_{i,j} (p(i,j)p^{\mathrm{del}}_a+(1-p(i,j))p^{\mathrm{add}}_a)$. 

\noindent Thus, we further have
\begin{equation}
\begin{aligned}
U_a
& =1 - \frac{\sum_{i,j} (p(i,j)p^{\mathrm{\mathrm{del}}}_a+(1-p(i,j))p^{{\mathrm{add}}}_a)}{\binom{N}{2}} \\
&= 1-(Rp^{del}_a +(1-R)p^{add}_a)
\end{aligned}
\end{equation}
Similarly, the local neighborhood utility for the auxiliary graph is 
\begin{equation}
U_u=1-(Rp^{del}_u +(1-R)p^{add}_u)
\end{equation} 
\end{idefinition}

Though the utility metric for structural data is application-dependent, our utility metric can provide a comprehensive understanding for utility performance by considering both the edge insertions and deletions, and incorporating the distance between the anonymized (auxiliary) graph and the conceptual underlying graph. Although our utility is one of the most straightforward definitions, to the best of our knowledge, it is still the first scientific work that theoretically analyzes the relationship between de-anonymization performance and the utility of the anonymized data.

Based on the local neighborhood utility in Definition~\ref{def_utility}, we theoretically analyze the de-anonymization capability of structure-based attacks and quantify the anonymized utility for successful de-anonymization. To improve readability, we defer the proof of Theorem \ref{maintheorem} to the Appendix. 

Theorem \ref{maintheorem} implies that as the number of nodes in the graphs $G_a$ and $G_u$ increase, the probability of successful deanonymization approaches $1$ when the four conditions (in Eqs.~\ref{t1},\ref{t4},\ref{t3},\ref{t2}) regarding graph density $R$, and the smallest and largest probabilities of the edges between nodes hold.
\begin{itheorem}\label{maintheorem}
For any $\sigma_k \neq \sigma_0$, where $k$ is the number of incorrectly-mapped nodes between $G_a$ and $G_u$, $\lim_{n\rightarrow \infty } Pr(\Phi_{\sigma_k}\geq \Phi_{\sigma_0})=1$ when the following conditions are satisfied.
\begin{equation}\label{t1}
 U_u+2lU_a > 1+2l-Rl
\end{equation} 
\begin{equation}\label{t4}
U_u +2(1-h)U_a > 1+2(1-h)-(1-h)(1-R)
\end{equation}
\begin{equation}\label{t3}
U_u +2l\frac{1-R}{R}U_a>1+2l\frac{1-R}{R}-l(1-R)
\end{equation}
\begin{equation}\label{t2}
U_u+2(1-h)\frac{R}{1-R}U_a > 1+2(1-h)\frac{R}{1-R}-R(1-h)
\end{equation} 
\end{itheorem}
From Theorem \ref{maintheorem}, we know that when the local neighborhood utility for the anonymized graph and the auxiliary graph satisfies the four conditions in Eqs.~\ref{t1},\ref{t4},\ref{t3},\ref{t2}, we can achieve successful de-anonymization \emph{from a statistical perspective}. The reason is that, the attacker can discover the correct mapping with high probability by choosing the mapping with the minimal Difference
of Common Neighbors (DCN), out of all the possible mappings between the anonymized graph and the auxiliary graph. To the best of our knowledge, this is the first work to quantify the relationship between anonymized utility and de-anonymization capability. It also essentially explains why structure-based de-anonymization attacks work. \\
\indent {The four conditions in Theorem \ref{maintheorem} can be reduced to one or two conditions under four types of graph density. Figure \ref{bound}(a) is the triangular utility region for $R<\min\{0.5,\frac{1-h}{1-h+l}\}$ (where the graph density $R$ is smaller than $0.5$ and $\frac{1-h}{1-h+l}$), which is only bounded by Eq. \ref{t1}. Figure \ref{bound}(b) is the quadrilateral utility region for $\min\{0.5,\frac{1-h}{1-h+l}\}\le R<0.5$ (where the graph density $R$ is larger than $\frac{1-h}{1-h+l}$ and smaller than $0.5$), which is bounded by Eq. \ref{t1} and Eq. \ref{t4}. Similarly, Figure \ref{bound}(c) is the triangular utility region for $0.5\le R<\max\{0.5,\frac{l}{1-h+l}\}$ (where the graph density $R$ is larger than $\frac{l}{1-h+l}$ and $0.5$), which is only bounded by Eq.~\ref{t2}. Figure \ref{bound}(d) is the quadrilateral utility region for $R\ge \max\{0.5,\frac{l}{1-h+l}\}$ (where the graph density $R$ is larger than $0.5$ and smaller than $\frac{l}{1-h+l}$), which is bounded by Eq. \ref{t3} and Eq. \ref{t2}. Therefore, we not only analytically explain why the structure-based de-anonymization works, but also theoretically provide the bound of anonymized utlity for sucessful de-anonymization. When the anonymized utility satisfies the conditions in Theorem \ref{maintheorem} (or locates within the green shaded utility regions shown in Figure \ref{bound}), successful de-anonymization is theoretically achievable. 

\subsection{Relation Between the Global Structure Utility and De-anonymization Capability}
In Definition~\ref{def_utility}, we consider a straightforward local neighborhood utility metric, which evaluates the distortion between the \emph{adjacency matrices} of the two graphs, i.e., $||{\bm A}_a-{\bm A}||_1$. However, the real-world data utility is application-oriented such that we need to consider a more general utility metric, to incorporate more aggregate information of the graph instead of just the \emph{adjacency matrix}. Motivated by the general utility distance in~\cite{Mittal2013NDSS,liu2016linkmirage}, we consider to utilize the $w$-th power of the transition probability matrix $T^w$, which is induced by the $w$-hop random walk on graph $G$, to define the global structure utility as follows:
\begin{idefinition}\label{def_utility_general} The global structure utility for the anonymized graph $G_a$ is defined as
\begin{equation}\label{global_utility}
U_{a(w)}=1-\frac{||{\bm T}_a^w-{\bm T}^{w}||_1}{2N}
\end{equation}
where ${\bm T}_a^w,{\bm T}^w$ are the $w$-th power of the transition probability matrix ${\bm T}_a,{\bm T}$, respectively. The denominator in Eq.~\ref{global_utility} is a normalization factor to guarantee $U_{a(w)}\in [0,1]$.
Similarly, the global structure utility for the auxiliary graph is 
\begin{equation}\label{general_utility}
U_{u(w)}=1-\frac{||{\bm T}_u^w-{\bm T}^w||_1}{2N}
\end{equation} 
\end{idefinition}

Our metric of global structure utility in Definition~\ref{def_utility_general} is intuitively reasonable for a broad class of real-world applications, and captures the $w$-hop random walks between the original graph $G$ and the anonymized graph $G_a$. We note that random walks are closely linked to structural properties of real-world data. For example, a lot of high-level social network based applications such as recommendation systems~\cite{andersen2008trust}, Sybil defenses~\cite{yu2008sybillimit} and anonymity systems~\cite{mittal2012pisces} directly perform random walks in their protocols. The parameter $w$ is application specific; for applications that require access to fine-grained community structure, such as recommendation systems~\cite{andersen2008trust}, the value of $w$ should be small. For other applications that utilize coarse and macro community structure of the data, such as Sybil defense mechanisms~\cite{yu2008sybillimit}, $w$ can be set to a larger value (typically around 10). Therefore, our global structure utility metric can quantify the utility performance of a perturbed graph for various real-world applications in a general and universal manner. \\
\indent
Based on this general utility metric, we further theoretically analyze the de-anonymization capability of structure-based attacks and quantify the anonymized utility for successful de-anonymization. To improve readability, we defer the proof of Theorem \ref{maintheorem_general} to the Appendix.

\begin{itheorem}\label{maintheorem_general}
For any $\sigma_k \neq \sigma_0$, where $k$ is the number of incorrectly-mapped nodes between $G_a$ and $G_u$, $\lim_{n\rightarrow \infty } Pr(\Phi_{\sigma_k}\geq \Phi_{\sigma_0})=1$ when the following conditions are satisfied:

\begin{scriptsize}
\begin{equation}\label{t1_general}
 U_{u(w)}+2lU_{a(w)} > 1+2l-\frac{wRl(N-1)}{2}
\end{equation} 
\begin{equation}\label{t4_general}
U_{u(w)} +2(1-h)U_{a(w)} > 1+2(1-h)-\frac{w(N-1)(1-h)(1-R)}{2}
\end{equation}
\begin{equation}\label{t3_general}
U_{u(w)} +2l\frac{1-R}{R}U_{a(w)}>1+2l\frac{1-R}{R}-\frac{wl(1-R)(N-1)}{2}
\end{equation}
\begin{equation}\label{t2_general}
U_{u(w)}+2(1-h)\frac{R}{1-R}U_{a(w)} > 1+2(1-h)\frac{R}{1-R}-\frac{wR(1-h)(N-1)}{2}
\end{equation}
\end{scriptsize}
\end{itheorem}

\indent Similar to Theorem~\ref{maintheorem}, when the global structure utility for the anonymized graph and the auxiliary graph satisfies all of the four conditions in Theorem \ref{maintheorem_general}, we can achieve successful de-anonymization from a statistical perspective. With rather high probability, the attacker can find out the correct mapping between the anonymized graph and the auxiliary graph, by choosing the mapping with the minimal DCN out of all the potential mappings.\\
\indent Furthermore, both Theorem \ref{maintheorem} and Theorem \ref{maintheorem_general} give meaningful guidelines for future designs of de-anonymization and anonymization schemes: 1) Since successful de-anonymization is theoretically achievable when the anonymized utility satisfies the conditions in Theorem \ref{maintheorem} (for the local neighborhood utility) and Theorem \ref{maintheorem_general} (for the global structure utility), the gap between the practical de-anonymization accuracy and the theoretically achievable performance can be utilized to evaluate the effectiveness of a real-world de-anonymization attack; 2) we can also leverage Theorem \ref{maintheorem} and Theorem \ref{maintheorem_general} for designing future secure data publishing to defend against de-anonymization attacks. For instance, a secure data publishing scheme should provide anonymized utility that locates out of the theoretical bound (green shaded region) in Figure \ref{bound} while enabling real-world applications. We will provide a practical analysis for such privacy and utility tradeoffs in Section~\ref{tradeoff}.

\begin{table}[!t]\scriptsize
\centering
\caption{De-anonymization \emph{Accuracy} of State-of-the-Art Approaches.}
\begin{tabular}{|c|c|c|c|}
\hline
{Datasets} & {$noise=0.05$} & {$noise=0.15$} & {$noise=0.25$}\\ \hline
{\cite{ji2014CCS}} & 0.95 & 0.81 & 0.73\\ \hline
\cite{nilizadeh2014CCS} & 0.83 & 0.74 & 0.68\\ \hline
\end{tabular}
\label{tradeoff_results}
\end{table}
\section{Practical Privacy and Utility Trade-off} \label{tradeoff}
In this section, we show how the theoretical analysis in Section~\ref{sec:quantification} can be utilized to evaluate the privacy risks of practical data publishing and the performance of practical de-anonymization attacks. To enable real-world applications without compromising the privacy of users, a secure data anonymization scheme should provide anonymized utility which does not locate within the utility region for perfect de-anonymization shown as the green shaded regions in Figure \ref{bound} (a)~(d). From a data publisher's point of view, we consider the worst-case attacker who has access to perfect auxiliary information, i.e., $noise_u=0$. Based on Theorem \ref{maintheorem}, we aim to quantify the amount of noise that is added to the anonymized data for achieving successful de-anonymization. After careful derivations, we know that when the $noise$ of the anonymized graph is less than 0.25 ({note that our derivation is from a statistical point of view instead of from the perspective of a concrete graph}), successful de-anonymization can be theoretically achieved (proof is deferred to the Appendix). Therefore, when the $noise$ added to the anonymized graph is less than 0.25, there would be a serious privacy breach since successful de-anonymization is theoretically achievable. Note that such a utility bound only conservatively provides the minimum noise that should be added to the anonymized data. Practically, we suggest a real-world data publisher to add more noise to protect the privacy of the data. Furthermore, such privacy-utility trade-off can be leveraged as a guide for designing new anonymization schemes. 

In addition, our derived theoretical analysis can also be utilized to evaluate the performance of existing de-anonymization attacks. We first implement our experiments on the Facebook dataset \cite{viswanath2009OSN} which contains $46,952$ nodes (i.e., users) connected by $876,993$ edges (i.e., social relationships). To evaluate the performance of existing de-anonymization attacks, we consider a popular perturbation method of Hay et al. in~\cite{hay2007anonymizing}, which applies a sequence of $r$ edge deletions followed by $r$ random edge insertions. A similar perturbation process has been utilized for the de-anonymization attacks in \cite{nilizadeh2014CCS}. Candidates for edge deletion are sampled uniformly at random from the space of the existing edges in graph $G$, while candidates for edge insertion are sampled uniformly at random from the space of edges that are not existing in $G$. Here, we define $noise$ (perturbations) as the extent of edge modification, i.e., the ratio of altered edges $r$ to the total number of edges, i.e., $noise=\frac{r}{M}$. Note that we add the same amount of $noise$ to the original graph of the {Facebook} dataset to obtain the anonymized graph and the auxiliary graph, respectively. Then, we apply the state-of-the-art de-anonymization attacks in \cite{ji2014CCS} and \cite{nilizadeh2014CCS} to de-anonymize the anonymized graph by leveraging the auxiliary graph. 

We utilize \emph{Accuracy} as an effective evaluation metric to measure the de-anonymization performance. \emph{Accuracy} is the ratio of the correctly de-anonymized nodes out of all the overlapped nodes between the anonymized graph and the auxiliary graph:
{
\begin{equation}\label{accuracy_metric}
\emph{Accuracy}=\frac{N_{cor}}{|V_a\cap V_u|},
\end{equation} }
where $N_{cor}$ is the number of correctly de-anonymized nodes. The \emph{Accuracy} of these de-anonymization attacks corresponding to different levels of $noise$ is shown in Table~\ref{tradeoff_results}.  

From Table~\ref{tradeoff_results}, we can see that the state-of-the-art de-anonymization attacks can only achieve less than $75\%$ de-anonymization accuracy when the $noise$ is 0.25, which demonstrates the ineffectiveness of previous work and the potential of developing more powerful de-anonymization attacks in the future. 

\section{\uppercase{Discussion}}
{\bf There is a clear trade-off between utility and privacy for data publishing.} In this work, we analytically quantify the relationships between the utility of anonymized data and the de-anonymization capability. Our quantification results show that privacy could be breached if the utility of anonymized data is high. Hence, striking the balance between utility and privacy for data publishing is important yet difficult - providing the high utility for real-world applications would decrease the data's resistance to de-anonymization attacks.

{\bf Suggestions for Secure Data Publishing.} Secure data publishing (sharing) is important for companies (e.g., online social network providers), governments and researchers. Here, we give several general guidelines: ($i$) Data owners should carefully evaluate the potential vulnerabilities of the data before publishing. For example, our quantification result in Section \ref{sec:quantification} can be utilized to evaluate the vulnerabilities of the structural data. ($ii$) Data owners should develop proper policies on data collections to defend against adversaries who aim to leverage auxiliary information to launch de-anonymization attacks. To mitigate such privacy threats, online social network providers, such as Facebook, Twitter, and Google+, should reasonably limit the access to users' social relationships. 
\vspace{-2em}

\section{\uppercase{Conclusion}}
In this paper, we first address several fundamental open problems in the structure-based de-anonymization research by quantifying the conditions for successful de-anonymization under a general graph model. Next, we analyze the capability of structure-based de-anonymization methods from a theoretical point of view. We further provided theoretical bounds of the anonymized utility for successful de-anonymization. Our analysis provides a theoretical foundation for structure-based de-anonymization attacks, and can serve as a guide for designing new de-anonymization/anonymization systems in practice. Future work can include studying our utility versus privacy trade-offs for more datasets, and designing more powerful anonymization/de-anonymization approaches.
\vspace{-2em}

\section*{\uppercase{Acknowledgements}}

\noindent This work was supported in part by NSF award numbers SaTC-1526493, CNS-1218817, CNS-1553437, CNS-1409415, CNS-1423139, CCF-1617286, and faculty research awards from Google, Cisco and Intel.

\vfill
\bibliographystyle{apalike}
{\small
\bibliography{bib}}

\begin{thebibliography}{}

\bibitem[Andersen et~al., 2008]{andersen2008trust}
Andersen, R., Borgs, C., Chayes, J., Feige, U., Flaxman, A., Kalai, A.,
  Mirrokni, V., and Tennenholtz, M. (2008).
\newblock Trust-based recommendation systems: an axiomatic approach.
\newblock In {\em WWW}.

\bibitem[Backstrom et~al., 2007]{backstrom2007WWW}
Backstrom, L., Dwork, C., and Kleinberg, J. (2007).
\newblock Wherefore art thou r3579x?: anonymized social networks, hidden
  patterns, and structural steganography.
\newblock In {\em WWW}.

\bibitem[Buccafurri et~al., 2015]{buccafurri2015discovering}
Buccafurri, F., Lax, G., Nocera, A., and Ursino, D. (2015).
\newblock Discovering missing me edges across social networks.
\newblock {\em Information Sciences}.

\bibitem[Dwork, 2006]{dwork2011differential}
Dwork, C. (2006).
\newblock Differential privacy.
\newblock In {\em Encyclopedia of Cryptography and Security}. Springer.

\bibitem[Erd{\H{o}}s and R{\'e}nyi, 1976]{erdHos:selected76}
Erd{\H{o}}s, P. and R{\'e}nyi, A. (1976).
\newblock On the evolution of random graphs.
\newblock {\em Selected Papers of Alfr{\'e}d R{\'e}nyi}.

\bibitem[Fabiana et~al., 2015]{fabiana2015anonymizing}
Fabiana, C., Garetto, M., and Leonardi, E. (2015).
\newblock De-anonymizing scale-free social networks by percolation graph
  matching.
\newblock In {\em INFOCOM}.

\bibitem[Hamming, 1950]{hamming1950belljournal}
Hamming, R.~W. (1950).
\newblock Error detecting and error correcting codes.
\newblock {\em Bell System technical journal}.

\bibitem[Hay et~al., 2008]{hay2008VLDB}
Hay, M., Miklau, G., Jensen, D., Towsley, D., and Weis, P. (2008).
\newblock Resisting structural re-identification in anonymized social networks.
\newblock {\em VLDB Endowment}.

\bibitem[Hay et~al., 2007]{hay2007anonymizing}
Hay, M., Miklau, G., Jensen, D., Weis, P., and Srivastava, S. (2007).
\newblock Anonymizing social networks.
\newblock {\em Computer Science Department Faculty Publication Series}.

\bibitem[Ji et~al., 2015a]{ji2015your}
Ji, S., Li, W., Gong, N.~Z., Mittal, P., and Beyah, R. (2015a).
\newblock On your social network de-anonymizablity: Quantification and large
  scale evaluation with seed knowledge.
\newblock In {\em NDSS}.

\bibitem[Ji et~al., 2015b]{ji2015secgraph}
Ji, S., Li, W., Mittal, P., Hu, X., and Beyah, R. (2015b).
\newblock Secgraph: A uniform and open-source evaluation system for graph data
  anonymization and de-anonymization.
\newblock In {\em USENIX Security Symposium}.

\bibitem[Ji et~al., 2014]{ji2014CCS}
Ji, S., Li, W., Srivatsa, M., and Beyah, R. (2014).
\newblock Structural data de-anonymization: Quantification, practice, and
  implications.
\newblock In {\em CCS}.

\bibitem[Korula and Lattanzi, 2014]{korula2014efficient}
Korula, N. and Lattanzi, S. (2014).
\newblock An efficient reconciliation algorithm for social networks.
\newblock {\em Proceedings of the VLDB Endowment}.

\bibitem[Liu et~al., 2016]{liu2016ddp}
Liu, C., Chakraborty, S., and Mittal, P. (2016).
\newblock Dependence makes you vulnerable: Differential privacy under dependent
  tuples.
\newblock In {\em NDSS}.

\bibitem[Liu and Mittal, 2016]{liu2016linkmirage}
Liu, C. and Mittal, P. (2016).
\newblock Linkmirage: Enabling privacy-preserving analytics on social
  relationships.
\newblock In {\em NDSS}.

\bibitem[Liu and Terzi, 2008]{liu2008SIGMOD}
Liu, K. and Terzi, E. (2008).
\newblock Towards identity anonymization on graphs.
\newblock In {\em SIGMOD}.

\bibitem[Machanavajjhala et~al., 2007]{machanavajjhala:TKDD07}
Machanavajjhala, A., Kifer, D., Gehrke, J., and Venkitasubramaniam, M. (2007).
\newblock l-diversity: Privacy beyond k-anonymity.
\newblock {\em ACM Transactions on Knowledge Discovery from Data}.

\bibitem[Mittal et~al., 2013a]{Mittal2013NDSS}
Mittal, P., Papamanthou, C., and Song, D. (2013a).
\newblock Preserving link privacy in social network based systems.
\newblock In {\em NDSS}.

\bibitem[Mittal et~al., 2013b]{mittal2012pisces}
Mittal, P., Wright, M., and Borisov, N. (2013b).
\newblock Pisces: Anonymous communication using social networks.
\newblock {\em NDSS}.

\bibitem[Narayanan and Shmatikov, 2008]{Narayanan2008SP}
Narayanan, A. and Shmatikov, V. (2008).
\newblock {\em Robust de-anonymization of large sparse datasets}.
\newblock IEEE S\&P.

\bibitem[Narayanan and Shmatikov, 2009]{Narayanan2009SP}
Narayanan, A. and Shmatikov, V. (2009).
\newblock {\em De-anonymizing social networks}.
\newblock IEEE S\&P.

\bibitem[Newman, 2010]{Newman2010oxford}
Newman, M. (2010).
\newblock {\em Networks: an introduction}.
\newblock Oxford University Press.

\bibitem[Newman, 2003]{Newman2003SIAM}
Newman, M.~E. (2003).
\newblock The structure and function of complex networks.
\newblock {\em SIAM review}.

\bibitem[Nilizadeh et~al., 2014]{nilizadeh2014CCS}
Nilizadeh, S., Kapadia, A., and Ahn, Y.-Y. (2014).
\newblock Community-enhanced de-anonymization of online social networks.
\newblock In {\em CCS}.

\bibitem[Pedarsani et~al., 2013]{pedarsani2013Allerton}
Pedarsani, P., Figueiredo, D.~R., and Grossglauser, M. (2013).
\newblock A bayesian method for matching two similar graphs without seeds.
\newblock In {\em Allerton}.

\bibitem[Pedarsani and Grossglauser, 2011]{Pedarsani2011SIGKDD}
Pedarsani, P. and Grossglauser, M. (2011).
\newblock On the privacy of anonymized networks.
\newblock In {\em SIGKDD}.

\bibitem[Pham et~al., 2013]{pham2013SIGMOD}
Pham, H., Shahabi, C., and Liu, Y. (2013).
\newblock Ebm: an entropy-based model to infer social strength from
  spatiotemporal data.
\newblock In {\em SIGMOD}.

\bibitem[Sala et~al., 2011]{sala:imc11}
Sala, A., Zhao, X., Wilson, C., Zheng, H., and Zhao, B.~Y. (2011).
\newblock Sharing graphs using differentially private graph models.
\newblock In {\em IMC}.

\bibitem[Sharad and Danezis, 2013]{sharad2013anonymizing}
Sharad, K. and Danezis, G. (2013).
\newblock De-anonymizing d4d datasets.
\newblock In {\em Workshop on Hot Topics in Privacy Enhancing Technologies}.

\bibitem[Sharad and Danezis, 2014]{sharad2014automated}
Sharad, K. and Danezis, G. (2014).
\newblock An automated social graph de-anonymization technique.
\newblock In {\em Proceedings of the 13th Workshop on Privacy in the Electronic
  Society}. ACM.

\bibitem[Srivatsa and Hicks, 2012]{Srivatsa2012CCS}
Srivatsa, M. and Hicks, M. (2012).
\newblock Deanonymizing mobility traces: Using social network as a
  side-channel.
\newblock In {\em CCS}.

\bibitem[Viswanath et~al., 2009]{viswanath2009OSN}
Viswanath, B., Mislove, A., Cha, M., and Gummadi, K.~P. (2009).
\newblock On the evolution of user interaction in facebook.
\newblock In {\em ACM workshop on Online social networks}.

\bibitem[Yu et~al., 2008]{yu2008sybillimit}
Yu, H., Gibbons, P.~B., Kaminsky, M., and Xiao, F. (2008).
\newblock Sybillimit: A near-optimal social network defense against sybil
  attacks.
\newblock In {\em IEEE S\&P}.

\end{thebibliography}
\section*{\uppercase{Appendix}}
\subsection{Proof of Theorem \ref{maintheorem}}
\emph{Proof Sketch:} First, we aim to derive $p^{{\mathrm{add}}}_{ua}(i,j)$ which is the projection process from $G_u$ to $G_a$ and $p^{\mathrm{del}}_{ua}(i,j)$ which is the deletion process from $G_u$ to $G_a$, to have $p^{{\mathrm{add}}}_{ua}(i,j)=  \frac{p^{{\mathrm{add}}}_a(1-p^{{\mathrm{add}}}_u)(1-p(i,j))+(1-p^{\mathrm{del}}_a)p^{\mathrm{del}}_up(i,j)}{(1-p^{{\mathrm{add}}}_u)(1-p(i,j))+p^{\mathrm{del}}_up(i,j)}$ and $p^{\mathrm{del}}_{ua}(i,j)= $ \\$\frac{(1-p^{{\mathrm{add}}}_a)p^{{\mathrm{add}}}_u(1-p(i,j))+p^{\mathrm{del}}_a(1-p^{\mathrm{del}}_u)p(i,j)}{p^{{\mathrm{add}}}_u(1-p(i,j))+(1-p^{\mathrm{del}}_u)p(i,j)}$.

Then, we want to prove $ p^{{\mathrm{add}}}_{ua}(i,j)<\frac{1}{2}$ and $p^{\mathrm{del}}_{ua}(i,j)<\frac{1}{2}$. It is easy to show that they are equivalent to $(1-p^{{\mathrm{add}}}_u)(1-2p^{{\mathrm{add}}}_a)(1-p(i,j)) > p^{\mathrm{del}}_u(1-2p^{\mathrm{del}}_a)p(i,j)\label{eq1}$ and $ p^{{\mathrm{add}}}_u(1-2p^{{\mathrm{add}}}_a)(1-p(i,j)) < (1-p^{\mathrm{del}}_u)(1-2p^{\mathrm{del}}_a)p(i,j) \label{eq2}$. From Eqs. \ref{t1},\ref{t2},\ref{t3},\ref{t4}, we have $ \frac{1}{2}>\max \{ p^{\mathrm{add}}_u, p^{\mathrm{del}}_u\}$. Similarly, we have $\frac{1}{2}> \max\{ p^{{\mathrm{add}}}_a,p^{{\mathrm{del}}}_a\}$. Now, we consider four different situations (a) $p^{\mathrm{del}}_u \geq p^{\mathrm{add}}_u$ and $p^{\mathrm{del}}_a \geq p^{\mathrm{add}}_a$, (b) $p^{\mathrm{del}}_u \geq p^{\mathrm{add}}_u$ and $p^{\mathrm{del}}_a \leq p^{\mathrm{add}}_a$, (c) $p^{\mathrm{del}}_u \leq p^{\mathrm{add}}_u$ and $p^{\mathrm{del}}_a \geq p^{\mathrm{add}}_a$, and (d) $p^{\mathrm{del}}_u \leq p^{\mathrm{add}}_u$ and $p^{\mathrm{del}}_a \leq p^{\mathrm{add}}_a$ to prove $ p^{{\mathrm{add}}}_{ua}(i,j)<\frac{1}{2}$ and $p^{\mathrm{del}}_{ua}(i,j)<\frac{1}{2}$. Under $\sigma^k$, let $V^k_u \subseteq V_u$ be the set of incorrectly de-anonymized nodes, $E^k_u = \{e_u(i,j) |i \in V^k_u $ or $j \in V^k_u \}$ be the set of all the possible edges adjacent to at least one user in $V^k_u$, $E^\tau_u = \{e_u(i,j) |i, j \in V^k_u, (i, j) \in E^k_u,$ and $(j, i) \in E^k_u \}$ be the set of all the possible edges corresponding to transposition mappings in $\sigma^k$, and $E_u = \{e_u(i,j) |1 \leq i \neq j \leq n\}$ be the set of all the possible links on $V$. Furthermore, define $m^k = |E^k_u|$ and $m^\tau = |E^\tau_u |$. Then, we have $|V^k_u| = k$, $m^k =\binom{k}{2}+ k(n - k)$, $m^\tau \leq \frac{k}{2}$ and $|E_u| =\binom{n}{2}$. 

Now, we quantify $\Phi^{\sigma_0}$ from a statistical perspective. We define $\Phi^{\sigma^k,E^\prime}$ as the DCN caused by the edges in the set $E^\prime$ under the mapping $\sigma^k$. Based on the definition of DCN, we obtain $\Phi^{\sigma^k} = \Phi^{\sigma^k,E_u \backslash E^k_u} + \Phi^{\sigma^k , E^k_u \backslash E^\tau_u} + \Phi^{\sigma^k,E^{\tau}_u}$ and $\Phi^{\sigma_0} = \Phi^{\sigma_0,E_u \backslash E^k_u} + \Phi^{\sigma_0 , E^k_u \backslash E^\tau_u} + \Phi^{\sigma_0,E^{\tau}_u}$. Since $ \Phi^{\sigma^k,E_u \backslash E^k_u} = \Phi^{\sigma_0,E_u \backslash E^k_u} $ and $ \Phi^{\sigma^k,E^{\tau}_u}= \Phi^{\sigma_0,E^{\tau}_u}$, we can obtain $Pr(\Phi^{\sigma^k}\geq \Phi^{\sigma_0}) = Pr(\Phi^{\sigma^k, E^k_u \backslash E^\tau_u}\geq \Phi^{\sigma_0,E^k_u\backslash E^\tau_u})$. Considering $\forall e_u(i,j) \in E^k_u \backslash E^\tau_u$ under $\sigma^k$, we know $\Phi^{\sigma^k, e_u(i,j)} \sim B(1, p(i,j)_u (p(\sigma_{k(i)},\sigma_{k(j)})_u \times p^{\mathrm{del}}_{ua}(i,j)+(1-p(\sigma_{k(i)},\sigma_{k(j)})_u)\times (1-p^{\mathrm{add}}_{ua}(i,j)))+ (1-p(i,j)_u)(p(\sigma_{k(i)},\sigma_{k(j)})_u\times (1-p^{\mathrm{del}}_{ua}(i,j))+(1-p(\sigma_{k(i)},\sigma_{k(j)})_u)\times p^{\mathrm{add}}_{ua}(i,j)))$. Similarly, under $\sigma_0$ $\forall e_u(i,j) \in E^k_u \backslash E^\tau_u$, we obtain $\Phi^{\sigma_0, e_u(i,j)} \sim B(1,p(i,j)_up^{\mathrm{del}}_{ua}+(1-p(i,j)_u)p^{\mathrm{add}}_{ua})$.

 Let $\lambda^{\sigma_0, e_u(i,j)}$ and $\lambda^{\sigma^k, e_u(i,j)}$ be the mean of $\Phi^{\sigma_0, e_u(i,j)}$ and $\Phi^{\sigma^k, e_u(i,j)}$, respectively. Then, we have $\lambda^{\sigma^k, e_u(i,j)}>p(i,j)_up^{\mathrm{del}}_{ua}+(1-p(i,j)_u)p^{\mathrm{add}}_{ua} = \lambda^{\sigma_0, e_u(i,j)}$
then $Pr(\Phi^{\sigma^k, e_u(i,j)} >\Phi^{\sigma_0, e_u(i,j)})> 1-2\exp \left(-\frac{(\lambda^{\sigma^k, e_u(i,j)}-\lambda^{\sigma_0, e_u(i,j)})^2}{8\lambda^{\sigma^k, e_u(i,j)}\lambda^{\sigma_0, e_u(i,j)}}\right)
= 1-$\\ $2 \exp \left(-f(p(i,j)_u,p(\sigma^k(i),\sigma^k(j))_u m_2 \right)$, where $f(.,.)$ is a function of $p(i,j)_u$ and $p(\sigma^k(i),\sigma^k(j))_u$. After further derivations, we obtain $\lim_{n\rightarrow \infty} Pr(\Phi^{\sigma^k, E^k_u \backslash E^\tau_u}\geq \Phi^{\sigma_0, E^k_u \backslash E^\tau_u}) = 1$ and have Theorem \ref{maintheorem} proved.\\
\subsection{Proof of Theorem \ref{maintheorem_general}}
We first relate the adjacency matrix ${\bm A}$ with the transition probability matrix ${\bm T}$ as ${\bm A} = {\bm \Lambda}{\bm T}$, where ${\bm \Lambda}$ is a diagonal matrix and ${\bm \Lambda}(i,i) = deg(i)$. Then we analyze the utility distance for the anonymized graph. When $w=1$, we can prove $||{\bm A}_a-{\bm A}||_1 = ||{\bm \Lambda}_a{\bm T}_a-{\bm \Lambda}{\bm T}||_1 = ||({\bm \Lambda}_a{\bm T}_a-{\bm \Lambda}_a{\bm T} + {\bm \Lambda}_a{\bm T} -{\bm \Lambda}{\bm T})||_1 \geq ||{\bm \Lambda}_a||_1||{\bm T}_a-{\bm T}||_1$. Since the element in the diagonal of ${\bm \Lambda}_a$ is greater than 1, we have $||{\bm A}_a-{\bm A}||_1\geq||{\bm T}_a-{\bm T}||_1$. Therefore, we can obtain $U_a \leq U_{a(w)}$. Similarly, we have $U_u \leq U_{u(w)}$. Incorporating these two inequalities into Eqs. \ref{t1},\ref{t4},\ref{t3},\ref{t2}, we have Theorem \ref{maintheorem_general} satisfied under $w=1$. Next, we consider $w\geq 1$. It is easy to prove that $||{\bm T}_a^w-{\bm T}^w||_1 \leq w||{\bm T}_a-{\bm T}||_1$ so $||{\bm T}_a^w-{\bm T}^w||_1 \leq w||{\bm A}_a-{\bm A}||_1$. Therefore, we have $U_a \leq wU_{a(w)} +1 -w$. Similarly, we also have $U_u \leq wU_{u(w)} +1 -w$ for the auxiliary graph. Incorporating these two inequalities into Eqs. \ref{t1},\ref{t4},\ref{t3},\ref{t2}, we have Theorem \ref{maintheorem_general} proved.

\subsection{Proof of Utility and Privacy Trade-off}}
For the anonymization method of Hay et al. in~\cite{hay2007anonymizing}, we have $P^{del}_a = {k_a}/{M_a}$ and $P^{add}_a = {k_a}/{(\binom{N}{2}-M_a)}$. Similarly, we have $P^{del}_u = {k_u}/{M_u}$ and $P^{add}_u={k_u}/({\binom{N}{2}-M_u})$. Based on our utility metric in Definition~\ref{def_utility}, we have $U_u =1-2R\times noise_u$ and $U_a=1-2R\times noise_a$. Considering the sparsity property in most real-world structural graphs \cite{Narayanan2008SP}, the utility condition for achieving successful de-anonymization is restricted by Eq.~\ref{t1}, which can be represented as $noise_u+l \times noise_a<\frac{l}{2}$. 
Consider the worst-case attacker who has access to perfect auxiliary information, i.e., $noise_u=0$. Therefore, we have $noise_a <0.25$.

\vfill
\end{document}